\documentclass[aps,prl,twocolumn,showpacs,preprintnumbers,nofootinbib,float,longbibliography,superscriptaddress,notitlepage]{revtex4-2}

\usepackage[utf8]{inputenc}
\usepackage{epsfig}
\usepackage{float,amsmath,amssymb}
\usepackage{graphicx,yfonts}
\usepackage{url}
\usepackage{bbold}
\usepackage{physics}
\usepackage{enumitem}

\usepackage[]{mdframed}
\newcommand{\trento}{T$\mathrel{\protect\raisebox{-2.1pt}{R}}$ENTo}

\usepackage[breaklinks,colorlinks,citecolor=citcolor,urlcolor=blue,linkcolor=lcolor]{hyperref}
\definecolor{lcolor}{rgb}{0.5,0,0}
\definecolor{citcolor}{rgb}{0,0.3,0.0}

\newcommand{\gege}{$^{76}$Ge+$^{76}$Ge}

\newcommand{\threej}[6]{\begingroup\setlength{\arraycolsep}{0.1em}\left(\hskip -\arraycolsep\begin{array}{ccc} #1 & #3 & #5 \\ #2 & #4 & #6 \end{array}\hskip -\arraycolsep\right)\endgroup}

\newcommand{\jhat}[1]{\hat{#1}}

\newcommand{\ylm}[3]{\ensuremath{Y_{#1}^{#2}{#3}}}

\begin{document}

\title{Revealing the harmonic structure of nuclear two-body correlations\\in high-energy heavy-ion collisions}

\author{Thomas Duguet}
\email{thomas.duguet@cea.fr}
\affiliation{IRFU, CEA, Universit\'e Paris-Saclay, 91191 Gif-sur-Yvette, France}

\author{Giuliano Giacalone}
\email{giuliano.giacalone@cern.ch}
\affiliation{Theoretical Physics Department, CERN, CH-1211 Gen\`eve 23, Switzerland}

\author{Sangyong Jeon}
\email{sangyong.jeon@mcgill.ca}
\affiliation{Department of Physics, McGill University, 3600 University street, Montreal, QC, Canada H3A 2T8}

\author{Alexander Tichai}
\email{alexander.tichai@tu-darmstadt.de}
\affiliation{Technische Universit\"at Darmstadt, Department of Physics, 64289 Darmstadt, Germany}
\affiliation{ExtreMe Matter Institute EMMI, GSI Helmholtzzentrum f\"ur Schwerionenforschung GmbH, 64291 Darmstadt, Germany}
\affiliation{Max-Planck-Institut f\"ur Kernphysik, Saupfercheckweg 1, 69117 Heidelberg, Germany}

\begin{abstract}
Smashing nuclei at ultrarelativistic speeds and analyzing the momentum distribution of outgoing debris provides a powerful method to probe the many-body properties of the incoming nuclear ground states. Within a perturbative description of initial-state fluctuations in the quark-gluon plasma, we express the measurement of anisotropic flow in ultra-central heavy-ion collisions as the quantum-mechanical average of a specific set of operators measuring the harmonic structure of the two-body azimuthal correlations among nucleons in the colliding states. These observables shed a new light on spatial correlations in atomic nuclei, while enabling us to test the complementary pictures of nuclear structure delivered by low- and high-energy experiments on the basis of state-of-the-art theoretical approaches rooted in quantum chromodynamics.
\end{abstract}

\preprint{CERN-TH-2025-056}

\maketitle

Atomic nuclei are among the most complex quantum systems observed in nature. They exhibit a rich spectrum of many-body phenomena, leaving measurable fingerprints on processes spanning subnuclear \cite{Mantysaari:2023qsq,STAR:2024wgy} to astrophysical \cite{1998PhR...294..167S} scales, including exotic interactions relevant to searches for physics beyond the Standard Model \cite{Klos2016_wimp,Hoferichter2019_wimp,Hu:2021awl,Agostini:2022zub,Door:2024qqz,Caputo:2024doz,Li:2025vdp}. Despite nearly a century of investigations, a clear understanding of many nuclear properties, even in their ground states, remains elusive. Recent years have however witnessed important steps forward on both the theoretical and the experimental forefronts. 

Breakthroughs in many-body methods and the access to unprecedented computing power have led to tremendous progress in the development of first-principles, \textit{i.e.} \textit{ab initio},  approaches to nuclear structure~\cite{Hergert:2020bxy,Hebeler2021PR_3NF,Ekstrom:2022yea}. Such first-principles description rely on the application of scalable many-body frameworks \cite{Hagen14RPP,Herg16PR,Gandolfi:2020pbj,Soma:2020xhv,Tichai2020review,Lee:2025req} on the basis of inter-nucleon interactions explicitly rooted into low-energy effective field theories of the strong force described via quantum chromodynamics (QCD)~\cite{Epelbaum:2008ga,Machleidt:2016rvv,Epelbaum:2019kcf,Piarulli:2019cqu,Hammer:2019poc}. Although systematic computations across the nuclear chart are still in their infancy~\cite{Yao18IMGCM,Miyagi2020PRC_MultiShellIMSRG,Frosini2021mrI,Frosini2021mrII,Frosini2021mrIII,Novario2020a,Stroberg2021,Hagen2022PCC,Porro2024_gcmII,Tichai2024_dmrgN50,Belley2024PRL_M0nuGe76}, \textit{ab initio} calculations have recently proven able to address both closed- and open-shell systems in unprecedented regimes of nuclear masses \cite{Morris2018PRL_Sn100,Hu:2021trw,Otsuka:2022bcf,Elhatisari:2022zrb,Miyagi2022PRC_NO2B,Tichai:2023epe,Hebeler2023jacno,Hu:2024pee,Frosini:2024ajq,Sun:2024iht,Arthuis2024arxiv_LowResForces,Miyagi2024_magmom}, which are expected to extend further in the near future. In this context, an understanding of many-body correlations in atomic nuclei rooted in the elementary description of the strong force is within reach. 

On the experimental front, traditional low-energy nuclear structure experiments are being complemented by high-energy experiments conducted at the Relativistic Heavy Ion Collider (RHIC) or the Large Hadron Collider (LHC). Due to the completely different time scales at play, this novel platform to study nuclear structure \cite{Jia:2022ozr} reveals more vividly the inter-nucleon correlations characterizing the ground state of the collided species \cite{Giacalone:2023hwk} through the analysis of the momentum distributions of hadrons emitted from the quark-gluon plasma (QGP \cite{Teaney:2009qa,Busza:2018rrf}). Analysis of these experiments via a rigid-rotor model of the colliding ions have already revealed key information regarding collective correlations in these nuclei in terms of intrinsic deformations, including quadrupole \cite{STAR:2015mki,ALICE:2018lao,ATLAS:2019dct,CMS:2019cyz,ALICE:2021gxt}, triaxial \cite{ATLAS:2022dov,STAR:2024wgy}, octupole \cite{STAR:2021mii}, and hexadecapole \cite{Ryssens:2023fkv,Xu:2024bdh} deformations. Effectively, high-energy colliders have the potential to shed a complementary view of the nucleon dynamics in nuclear ground states.

However, while a wide array of more or less sophisticated nuclear models have been used as input to heavy-ion collision simulations that have subsequently proven successful in phenomenological applications~\cite{Giacalone:2017dud,Lim:2018huo,Rybczynski:2019adt,Schenke:2020mbo,Summerfield:2021oex,Bally:2021qys,Xu:2021uar,Nijs:2021clz,Zhang:2021kxj,Zhao:2022uhl,Giacalone:2023cet,Fortier:2023xxy,Giacalone:2024luz,Zhang:2024vkh,Fortier:2024yxs,Giacalone:2024ixe,Zhao:2024feh,Mantysaari:2024uwn,Ambrus:2024hks,Ambrus:2024eqa,Lu:2025cni,Mantysaari:2025tcg}, a deep understanding of the many-body correlations underlying the observables measured at colliders is still lacking. This situation can be improved via two complementary approaches:
\begin{enumerate}
\item Understanding on a more rigorous ground the character of the measured observables,
\item Performing a systematic analysis of such observables based on state-of-the-art \textit{ab initio} nuclear-structure calculations.
\end{enumerate}

In this Letter, we take a major step regarding point 1) in view of addressing point 2) on a solid footing in the future. We show that measurements of anisotropic flow in the final states of heavy-ion collisions can be formulated in terms of the quantum-mechanical average of a new class of two-body self-adjoint operators in the colliding ground states, giving access to the harmonic decomposition of their associated two-body density. 

To reach this goal, the route proceeds via a simple theoretical picture of initial-state fluctuations in heavy-ion collisions. As in most phenomenological approaches, a nucleus at high energy is associated with a density of matter in the plane transverse to the beam direction, or \textit{thickness} function,
\begin{equation}
\label{eq:thickness}
t({\bf x}) \equiv \sum_{i=1}^A g({\bf x}-{\bf r}_{\perp i}) \, ,
\end{equation}
where $A$ is the nuclear mass number, ${\bf x}=(x,y)$ denotes a position vector in the transverse plane with its origin located at the center-of-mass of the nucleus (${\bf x}=0$), while ${\bf r}_{\perp i}$ is the transverse position of the $i$th nucleon. The function $g({\bf x})$ parametrizes the two-dimensional shape of a nucleon as seen by a high-energy probe. 

Equation~(\ref{eq:thickness}) describes an individual nucleus in an individual collision. The elements of randomness are the positions of the nucleons in the transverse plane, ${\bf r}_{\perp i}$. In our understanding of high-energy collisions, the process resolves the instantaneous structure of the colliding objects down to scales dictated by the momentum transfer, which is on the order of 1 GeV or higher.  Therefore, the nucleon positions are \textit{frozen} while the two nuclei cross each other \cite{Miller:2007ri,Alver:2010gr}. If the ground state is characterized by the $A$-body wave-function (omitting spin and isospin)
     \begin{equation}
   \Psi({\bf r}_1, \ldots, {\bf r}_A) \equiv \langle {\bf r}_1, \ldots, {\bf r}_A | \Psi \rangle  \, , 
 \end{equation}
the probability to find nucleons at coordinates ${\bf r}_i=({\bf r}_{\perp i}, r_{z i})$ is given by $| \langle {\bf r}_1, \ldots, {\bf r}_A | \Psi \rangle |^2$. Thus, the nucleon positions sampled according to this probability distribution reflect up to $A$-body correlations among them\footnote{High-energy modifications of the nucleon structure, \textit{e.g.} the fact that different nucleons can have different gluon content, or the possibility of non-linear interactions among nucleons in Eq.~(\ref{eq:thickness}) owing to nuclear parton distribution functions \cite{Klasen:2023uqj} or gluon saturation \cite{Morreale:2021pnn} are not discussed here. These effects pertain to the short-scale partonic structure of the colliding ions and do not affect the conclusions of our discussion.}.

Given the thickness functions $t({\bf x})$ and $t^\prime({\bf x})$ of two colliding nuclei, generic data-driven arguments are employed to move to the initial entropy density of the subsequently formed QGP. First, the interactions of partons from the colliding nuclei are semi-hard processes, which in the transverse plane are localized over small fractions of the size of a nucleon. Therefore, the entropy deposition at a given location ${\bf x}$ can only depend on the values of the two thickness functions in the immediate vicinity of such a point, \textit{i.e.} $t({\bf x})$ and $t^\prime ({\bf x})$. Second, the QGP is initially so hot that it can be roughly viewed as an ideal gas of quarks and gluons, where the total entropy determines the particle number \cite{Ollitrault:2007du}. As the evolution of the QGP is nearly isentropic, this implies that the final-state multiplicity is dictated by the initial-state entropy. In turn, in the limit of ultra-central collisions (small impact parameters) it is observed experimentally that the final-state multiplicity scales with the mass number of the colliding nuclei \cite{ALICE:2018cpu}. Combining these points, the entropy density at midrapidity is of the form
\begin{equation}
s({\bf x})   = \mathcal{F} \left [ t({\bf x}), t^\prime ({\bf x}) \right ],  \hspace{15pt} \int_{\bf x} s({\bf x}) \propto A \, ,
\end{equation}
where $\mathcal{F}$ is some model-dependent function. The above reasoning underlies the popular \trento{} Ansatz for the entropy density of heavy-ion collisions \cite{Moreland:2014oya} and is supported by the results of a Bayesian analysis of LHC data with more generic initial conditions \cite{Nijs:2023yab}.

Next, the notion of \textit{anisotropic flow} of hadrons \cite{Ollitrault:2023wjk} is introduced to connect properties of the initial entropy density, $s({\bf x})$, to measurable quantities in the final states of the collisions. In a collision, hadrons are emitted at midrapidity with some azimuthal distribution
\begin{equation}
\label{eq:vn}
    \frac{dN}{d\phi} \propto 1 + 2 \, v_2 \cos(2( \phi-\phi_2 )) + 2 \, v_3 \cos(3 (\phi-\phi_3)) + \ldots \, .
\end{equation}
The set of complex Fourier coefficients $V_n=v_ne^{in\phi_n}, n\geq 2,$ defines the anisotropic flow of the system.  Because of the hydrodynamic nature of the QGP, the anisotropy coefficients in momentum space, $V_n$, are determined by the anisotropies in coordinate space, $\mathcal{E}_n$, characterizing the large-scale structure of the initial entropy density field \cite{Teaney:2010vd}. These quantities are defined by [with ${\bf x}=(|{\bf x}|,\phi_x)$]
\begin{equation}
\label{eq:newEn}
    \mathcal{E}_n \equiv - \frac{ \int_{\bf x} |{\bf x}|^n e^{in\phi_x} s({\bf x}) }{\int_{{\bf x}} |{\bf x}|^n s({\bf x})} , \hspace{20pt} \varepsilon_n = |\mathcal{E}_n| \ .
\end{equation}

Given the probabilistic nature of the nucleon distributions in the colliding nuclei, the field $s({\bf x})$ and consequently the value of $\mathcal{E}_n$ fluctuate event by event. Hydrodynamic studies show, then, that $v_n$ in the final state is linearly correlated with the value of $\varepsilon_n$ in the initial state: if we observe a fluctuation in the value of $\varepsilon_n$, a similar variation will be also observed in the value of $v_n$, \textit{i.e.}, $v_n \propto \varepsilon_n$ \cite{Sousa:2024msh}. Following Ref.~\cite{Blaizot:2014nia}, a background-fluctuations splitting is thus introduced according to 
\begin{equation}
    s({\bf x}) = \langle s({\bf x}) \rangle  + \delta s({\bf x}),\hspace{25pt}\langle \delta s({\bf x}) \rangle =0 \, , \label{fluctuations}
\end{equation}
where the brackets denote a statistical average over ultra-central collision events. Inserting Eq.~\eqref{fluctuations} into the expression of $\mathcal{E}_n$ in Eq.~(\ref{eq:newEn}), expanding in powers of $\delta s$ and keeping terms up to second order, one obtains
\begin{equation}
\label{eq:linEn}
\langle v_n^2 \rangle \propto  \langle \varepsilon_n^2 \rangle =    \frac{ \int_{{\bf x}, {\bf y}} |{\bf x}|^n~|{\bf y}|^n~e^{in(\phi_x-\phi_y)} C_2({\bf x}, {\bf y}) } { \left ( \int_{\bf x} |{\bf x}|^n ~C_1({\bf x}) \right )^{2} } \, ,
\end{equation}
where the 1-point and connected 2-point functions of the entropy density field have been introduced
\begin{align}
\label{eq:C2}
    \nonumber C_1({\bf x}) &\equiv  \langle s({\bf x}) \rangle, \\
    C_2({\bf x}, {\bf y}) &\equiv  \langle s({\bf x})s({\bf y}) \rangle- C_1({\bf x})C_1({\bf y}).
\end{align}
As shown in Ref.~\cite{Giacalone:2023hwk}, Eq.~(\ref{eq:linEn}) is accurate at the percent level for collisions of nuclei with $A\sim 100$. This completes our leading-order picture of initial-state and spatial-anisotropy fluctuations in heavy-ion collisions. The variations of a quantity that is experimentally accessible, the flow coefficient $v_n$, are determined by the variations of an initial-state quantity, $\varepsilon_n$, whose fluctuations are in turn related to the correlation functions of the entropy density field, $\langle s({\bf x}) \rangle$ and $\langle s({\bf x}) s({\bf y}) \rangle$, via a perturbative expansion.

One important link is now added to the above scheme by connecting the fluctuations of the entropy density to the many-body wave functions of the colliding nuclei. In practice, a model of $s({\bf x})$ satisfying the generic requirements outlined above is needed. The simplest choice, amenable to an analytical treatment, corresponds to an arithmetic average\footnote{As demonstrated in the Supplemental Material (SM), our conclusions do not depend on the use of this particular model.}
\begin{equation}
\label{eq:save}
    s({\bf x}) = s_0 \frac{t({\bf x})+t^\prime({\bf x})}{2} \, ,
\end{equation}
where $s_0$ is a normalization typically inferred from experimental data. When the two colliding nuclei are nearly identical, one has $s({\bf x}) \propto t({\bf x})$.  The entropy density thus becomes akin to that discussed in so-called independent source model calculations, which have been studied at large in heavy-ion collisions \cite{Bhalerao:2006tp,Bhalerao:2011yg,Bhalerao:2011bp,Floerchinger:2013vua,Blaizot:2014nia,Gronqvist:2016hym,Giacalone:2020lbm,Borghini:2024ekn}. In our case,  $A$ sources determine the scaling of the total entropy with the size of the colliding nuclei but their \textit{independent} character is abandoned given that their density in coordinate space is identified with the ground-state probability $| \langle {\bf r}_1, \ldots, {\bf r}_A | \Psi \rangle |^2$. This has a profound conceptual implication given that it allows one to commute the statistical average of the entropy density field into a quantum mechanical expectation in the nuclear ground state. As shown in the SM, Eq.~(\ref{eq:save}) leads to
 \begin{align}
 \label{eq:C12}
  \nonumber   C_1({\bf x}) &= s_0\,A \int_{{\bf r}_{1}} \rho^{(1)}({\bf r}_{1}) g({\bf x}-{\bf r}_{1\perp}), \\
\nonumber C_2({\bf x}, {\bf y})&= \frac{s_0^2}{2} \biggl [  A \int_{{\bf r}_{1}} \rho^{(1)} ({\bf r}) g({\bf x}-{\bf r}_{1\perp}) g({\bf y}-{\bf r}_{1\perp})  \\ 
\nonumber &\hspace{-25pt} + (A^2-A) \int_{{\bf r}_{1}, {\bf r}_{2}} \rho^{(2)} ({\bf r}_{1}, {\bf r}_{2}) g({\bf x}-{\bf r}_{1\perp}) g({\bf y}-{\bf r}_{2\perp}) \\
& \hspace{-25pt} - A^2 \int_{{\bf r}_1} \rho^{(1)}({\bf r}_1) g({\bf x}-{\bf r}_{1\perp})\int_{{\bf r}_2} \rho^{(1)}({\bf r}_2) g({\bf y}-{\bf r}_{2\perp}) 
\biggr ] \, ,
 \end{align}
where $\rho^{(1)}$ and $\rho^{(2)}$ denote the local one- and two-body ground-state densities, respectively, defined through
\begin{equation}
\rho^{(n)}({\bf r}_{1},\ldots,{\bf r}_{n})  \equiv \int_{{\bf r}_{n+1},\ldots,{\bf r}_A} | \Psi({\bf r}_{1},\ldots,{\bf r}_{A}) |^2 \, .
\end{equation}
As shown in the SM, inserting Eqs.~(\ref{eq:C12}) into Eq.~(\ref{eq:linEn}) and considering that the nucleon form factor $g({\bf x})$ is a highly localized function that can be approximated by a Dirac delta, we arrive at the following key expression
\begin{align}
\label{eq:Enquasifinal}
        \nonumber\langle \varepsilon_n^2 \rangle =& \frac{1}{2A^{2}} \frac{1}{\left ( \int_{{\bf r}} \rho^{(1)} ({\bf r} ) \hat{R}_{n} ({\bf r}) \right )^2 } \biggl [ A \int_{{\bf r}} \rho^{(1)}({\bf r}) \hat{R}_{2n} ({\bf r}) \\
        &+ (A^2-A) \int_{{\bf r}_{1},{\bf r}_{2}} \rho^{(2)}({\bf r}_{1},{\bf r}_{2 }) \hat{\mathcal{E}}_{n} ({\bf r}_{1},{\bf r}_2)  \biggr]  \\
        &= \frac{1}{2A} \frac{1}{  \langle \hat{R}_{n} \rangle^2  } \left [ \langle \hat{R}_{2n} \rangle + (A-1) \langle \hat {{\mathcal{E}}}_{n} \rangle \right ] \, , \nonumber
\end{align}
involving the ground-state expectation value of the one- and two-body operators 
\begin{subequations}
\begin{align}
    \hat{R}_{n} ({\bf r}) &\equiv r_{1\perp}^n = (r_x^2 + r_y^2)^{n/2} \, , \\
    \hat{\mathcal{E}}_{n} ({\bf r}_{1},{\bf r}_2) & \equiv   (r_{1x}+ir_{1y})^n(r_{2x}-ir_{2y})^{n} \nonumber \\
    &= c^{-1}_n  \,  r_1^n \, \ylm{n}{n}{(\Omega_1)}  \, c^{-1}_{-n} \, r_2^n \, \ylm{n}{-n}{(\Omega_2)} \nonumber \\
    &\equiv \hat{F}_{n} ({\bf r}_1) \hat{F}_{-n} ({\bf r}_2) \, ,
\end{align}
\end{subequations}
where the derivation of the last equality involving the (complex) spherical harmonic in the maximal projection
\begin{equation}
    \ylm{n}{\pm n}{(x,y,z)} \equiv c_{\pm n} \frac{ (x \pm i y)^n}{(x^2+y^2 +z^2)^{n/2}} \, ,
\end{equation}
is provided in the SM. 

The mean squared anisotropy $\langle \varepsilon_n^2 \rangle$ is thus expressed in terms of the ground-state expectation of a one-body transverse radius operator $\hat{R}_{n} ({\bf r})$ and a two-body \textit{eccentricity operator} $\hat{\mathcal{E}}_{n} ({\bf r}_{1},{\bf r}_2)$. The latter probes the azimuthal anisotropy of the local two-body density in the transverse plane. The appearance of the maximal projections of the spherical harmonics is indeed due to the Lorentz contraction of the three-dimensional nuclear state. The operator $\hat{\mathcal{E}}_{2} ({\bf r}_{1},{\bf r}_2)$ bears some resemblance with the two-body Kumar quadrupole operator~\cite{Kumar:1972zza}. However, while no low-energy experiment has been identified to directly measure Kumar's observables in nuclear ground states,
the novel set of eccentricity operators can be measured through the elliptic flow in symmetric\footnote{An asymmetric configuration where $t({\bf x})$ and $t^\prime ({\bf x})$ label two different species in Eq.~(\ref{eq:save}) makes the present analysis more complicated, as the quantities in Eq.~(\ref{eq:C12}) become sums of contributions from different nuclei. However, it may be possible that certain combinations of observables coming from selected collision systems (\textit{e.g.} isobars \cite{Jia:2021oyt}) may offer an increased sensitivity to some target quantity. This will be studied in a future work.} nucleus-nucleus collisions at high-energy. Note that, much as higher-order Kumar parameters enable one to extract additional information about nuclear collectivity, it will also be possible to derive an expression akin to Eq.~(\ref{eq:Enquasifinal}) for other types of correlators that will probe the local three-body density \cite{Giacalone:2023hwk}.

In the uncorrelated limit where $\rho^{(2)}({\bf r}_1, {\bf r}_2) = \rho^{(1)}({\bf r}_1) \rho^{(1)}({\bf r}_2)$, the second term in the rhs of Eq.~(\ref{eq:Enquasifinal}) vanishes due to the spherical symmetry of the local one-body density of $J=0$ even-even ground states. One is left with the first term associated with the transverse radius operator corresponding to the independent-source result proportional to $1/A$ originally derived in Ref.~\cite{Bhalerao:2006tp}. 

In reality, any ground state displays non-zero two-body correlations that are thus captured by the second term in the rhs of Eq.~(\ref{eq:Enquasifinal})  via the two-body operator $\hat{\mathcal{E}}_{n} ({\bf r}_{1},{\bf r}_2)$. For example, quadrupole correlations among nucleons {\it in the ground state} are directly accessed in the transverse plane via $\left \langle v_2^2 \right \rangle$, \textit{i.e.} the larger $\left \langle v_2^2 \right \rangle$, the larger these correlations. This is, in essence, our main result.

While these novel observables can quantify specific two-body spatial correlations in atomic nuclei, they cannot, per se, characterize the inner mechanisms that generate them. This can only be achieved {\it within} a given theoretical scheme, in a way that depends on such a scheme. Our ambition is eventually to combine the measurement of these collider observables for a portfolio of nuclei and to analyze whether and how a given {\it ab initio} theoretical scheme reproduces them. For example, expansion many-body methods proceeding through the breaking and restoration of rotational symmetry along with the resummation of missing dynamical correlations~\cite{Frosini2021mrI,Frosini2021mrII,Frosini2021mrIII,Hagen2022PCC}, can quantify the extent to which, \textit{e.g.}, quadrupole correlations measured via the fluctuations of the mean squared anisotropy are due to static quadrupole deformation, quadrupole shape fluctuations and residual non-collective two-body correlations. This is a long-term research program that the present work seeks to motivate.

For now, let us proceed through such an analysis based on an elementary nuclear model that has proven useful in heavy-ion collision simulations, \textit{i.e.} the textbook rigid-rotor model where the incoming nuclei are modeled as a (potentially) deformed intrinsic shapes rotating in all possible ways~\cite{Jia:2021tzt,Jia:2021qyu}. More specifically, a colliding nucleus is treated as a batch of nucleons independently sampled from a deformed intrinsic density that has, in each collision, a random orientation in space. The model assumes, by construction, that spatial correlations in the nucleus can be entirely mocked up via the rotation of nucleons collectively aligned along a preferred direction in the intrinsic frame, \textit{i.e.} the potentially deformed shape is rigid and there is no concept of shape fluctuations or non-collective correlations. The deformations of the associated intrinsic density distribution are quantified via standard Bohr multipole parameters, $\beta_n$. Using such a nuclear model, one obtains~\cite{Giacalone:2021udy}
\begin{equation}
    \langle \varepsilon_n^2 \rangle = a_n + b_n \beta_n^2 \, , \label{rotor}
\end{equation}
where $a_n$ and $b_n$ are positive coefficients. The first and second terms in the right-hand side of Eq.~(\ref{rotor}) are naturally identified with the first and second terms in Eq.~(\ref{eq:Enquasifinal}), respectively. Within this model, one can conclude that the operator $\hat{\mathcal{E}}_n$ quantifies the (square of the) effective deformation of the intrinsic rotor\footnote{Modulo some prefactors, Eq.~(\ref{rotor}) demonstrates that, {\it within the rotor model}, the lab-frame expectation value of $\hat{F}_{n} ({\bf r}_1) \hat{F}_{-n} ({\bf r}_2)$ ($\beta_n^2$) is equal to the square of the expectation value of $\hat{F}_{n} ({\bf r})$ (or $\hat{F}_{-n} ({\bf r})$) in the intrinsic frame ($\beta_n$). This reflects the rigid/static character of the effective deformation in this model. This would typically not be the case in other theoretical approaches where two-body correlations are (notably but not only) impacted by the fact that the shape deformation, indeed apparent in the method, does not take a definite value but fluctuates.}. 

To illustrate more explicitly how the novel observables are interpreted within the rigid rotor model, the intrinsic density is taken as a simple Gaussian function 
\begin{equation}
    \rho(r,\Omega) \equiv \frac{1}{(2\pi)^{3/2} R^3} \exp \left ( -\frac{r^2}{2 R(\Omega)^2}\right ) \, ,
\end{equation}
with the nuclear surface expanded in quadrupole and octupole modes,
\begin{equation}
    R (\Omega) \equiv R \left [ 1 + \beta_{20} \ylm{2}{0}{(\Omega)}  + \beta_{30} \ylm{3}{0}{(\Omega)}    \right ] \, .
\end{equation}
From a Taylor expansion truncated to first order in the deformation parameters $\beta_{20}$ and $\beta_{30}$, the evaluation of Eq.~(\ref{eq:Enquasifinal}) for $n=2,3$ leads to
\begin{equation}
\label{eq:Engauss} 
      \left \langle \varepsilon_2^2 \right \rangle = \frac{1}{A}  +  \frac{3}{4\pi} \beta_{20}^2 ,     \hspace{15pt}    \left \langle \varepsilon_3^2 \right \rangle = \frac{16}{3\pi \, A}  +  \frac{2048}{245\pi^3} \beta_{30}^2 \,. 
\end{equation}
In each case, the first term proportional to $1/A$ corresponds to that derived in independent-source calculations with a Gaussian density of sources (see \textit{e.g.} \cite{Gronqvist:2016hym}). The second term delivers the correction induced by two-body correlations through intrinsic axial deformation parameters with prefactors matching exactly those obtained in Ref.~\cite{Jia:2021tzt} based on a liquid-drop-like model rotating in space. More sophisticated  calculations are reported in the SM, showing that the correction to $\langle \varepsilon_2^2 \rangle$ induced by the quadrupole deformation of the nucleus is robust and independent of the precise modeling of the collisions. 

Because correlations impact the two-body density in any nucleus, the rigid-rotor model is {\it bound} to send back a non-zero effective deformation for all nuclei through the second term of Eq.~(\ref{rotor}). As a matter of fact, this model has proven very successful in the analysis of heavy-ion collisions even for nuclei that do not show any rotational character, such as $^{197}$Au, or the isobars $^{96}$Ru and $^{96}$Zr. The case of $^{96}$Zr is especially interesting: data on elliptic and triangular flow in isobar collisions at RHIC appear to be consistent with the picture of an intrinsic rotor with a large octupole deformation parameter $\beta_3$ \cite{Zhang:2021kxj,Nijs:2021clz,Xu:2021uar,Zhao:2022uhl} extracted from Eq.~(\ref{eq:Engauss}), even though this is at odds with the transitional nature of this isotope. The effectiveness of such a type of modeling suggests that associating a nucleus with a rigid intrinsically deformed density rotating in space enables one to obtain azimuthal correlations consistent with the measured $\langle \hat{\mathcal{E}}_n \rangle$ values. Interestingly, the {\it picture} extracted from other pertinent\footnote{A theoretical scheme is pertinent if it reproduces quantitatively the value and the evolution of the {\it observables} of interest (among others) across a given set of nuclei. This must indeed be demonstrated and constitutes a prerequisite to proceed to the analysis of the mechanisms at play in such a model to reproduce the observables in question.} theoretical schemes is likely to be different even though the observables may be equally well reproduced. Still, it may be possible, \textit{i.e.} in {\it ab initio} projected generator coordinate method calculations explicitly including shape fluctuations~\cite{Giacalone:2024ixe,Giacalone:2024luz},  to successfully map the results of a different theoretical framework back onto a rigid-rotor picture, and extract an effective rigid deformation consistent with anisotropic flow measurements.

The above example illustrates how measurements of anisotropic flow in heavy-ion collisions, aimed at quantifying two-body correlations in nuclei, can help elucidate the inner mechanisms that generate such correlations in a given theoretical scheme. This is complementary to low-energy observables such as, \textit{e.g.}, quadrupole transition strengths from the ground state to the first $2^+$ excited state, $B(E2; 0^+ \rightarrow 2^+)$, that are traditionally used to characterize collectivity in nuclei. To motivate future experimental endeavors, our ambition is to analyze the novel observables within the frame of {\it ab initio} expansion many-body methods for doubly magic, singly magic, and doubly open-shell nuclei where the mechanisms building up two-body correlations are expected to vary considerably. Particularly interesting candidates for such an endeavor are $^{16}$O and $^{20}$Ne, which are also relevant for current and upcoming ion runs at colliders \cite{Huang:2023viw,Mariani:2024jzp,AlemanyFernandez:2025ixd}. Furthermore, the observables introduced in this paper should be evaluated while performing a global sensitivity analysis of the parameters of the nuclear interaction \cite{Ekstrom:2019lss,Ekstrom:2023nhc,Belley:2024zvt,Zhang:2024tzr}, to rigorously quantify how anisotropic flow measurements at colliders constrain the low-energy constants of nuclear forces based on chiral effective field theory. 

The authors thank Benjamin Bally, Jean-Paul Blaizot, Andreas Ekstr\"om, Mikael Frosini, Charles Gale, Gaute Hagen, Jiangyong Jia, Matt Luzum, Andrea Porro, Jean-Yves Ollitrault, Achim Schwenk, and Chunjian Zhang for useful discussions. We thank Wouter Ryssens for providing us with information on the structure of $^{76}$Ge. S.~J. and A.~T. acknowledge the hospitality of the Theoretical Physics Department of CERN, where this work was initiated.
This work was supported in part by the European Research Council (ERC) under the European Union's Horizon 2020 research and innovation programme (Grant Agreement No.~101162059).
S.~J acknowledges the support of the Natural Sciences and Engineering Research Council of Canada (NSERC) [SAPIN-2024-00026].

\section{Supplemental material}

\subsection{1- and 2-point functions of the entropy density}

We consider symmetric nucleus-nucleus collisions at zero impact parameter. We consider that all the nucleons from the incoming nuclei participate in the collisions. The entropy density at transverse position ${\bf x}$ is given by 
\begin{equation}
    s({\bf x}) = s_0 \frac{t({\bf x})+t^\prime({\bf x})}{2} \, ,
\end{equation}
where $s_0$ fixes the normalization, while the thickness functions are defined by
\begin{equation}
\label{eq:thickness}
t({\bf x}) \equiv \sum_{i=1}^A g({\bf x}-{\bf r}_{\perp i}) \, ,
\end{equation}
where $A$ is the nuclear mass number,  ${\bf r}_{\perp i}$ is the transverse position of the $i$th nucleon, whose original coordinates are ${\bf r}=({\bf r}_\perp, r_z)$. The function $g({\bf x})$ is the two-dimensional form factor of the nucleon at high energy. We now evaluate the first two cumulants of the entropy density field. Similar derivations can be found in, \textit{e.g.}, Refs.~\cite{Blaizot:2014wba,Giacalone:2023hwk}.

\subsubsection{One-point function}

The first is the local average, $\langle s({\bf x})\rangle$. As we collide the same nuclear species, and as all nucleons are treated equally, the one-point average is given by
\begin{align}
\nonumber    \langle s({\bf x})\rangle &= \frac{s_0}{2} \left ( \langle t({\bf x}) \rangle +  \langle t^\prime({\bf x}) \rangle \right ) \\
\nonumber    & = \frac{s_0}{2} \left ( \left \langle \sum_{i=1}^A g({\bf x}-{\bf r}_{\perp i}) \right \rangle +  \left \langle \sum_{i=1}^A g({\bf x}-{\bf r}_{\perp i}) \right \rangle \right )  \\
\nonumber    & = s_0  \sum_{i=1}^A \left \langle g({\bf x}-{\bf r}_{\perp i}) \right \rangle \\
\label{eq:aveSappe}    &= s_0 \, A \, \langle g({\bf x}-{\bf r}_{\perp i}) \rangle. 
\end{align}
 The angular brackets in these expressions denote averages over events, which correspond to the expectation values of $n$-body operators, say $ \hat{\mathcal{O}}({\bf r}_1, \ldots, {\bf r}_n )$, which are functions of $n$ random positions. In general, these expectation values are of the following form: 
\begin{equation}
    \langle \hat{\mathcal{O}}({\bf r}_1, \ldots, {\bf r}_n )\rangle = \frac{\int_{{\bf r}_1, \ldots , {\bf r}_n} \mathcal{O}({\bf r}_1, \ldots, {\bf r}_n) \rho^{(n)}({\bf r}_1, \ldots, {\bf r}_n)}{\int_{{\bf r}_1, \ldots , {\bf r}_n} \rho^{(n)}({\bf r}_1, \ldots, {\bf r}_n)}.
\end{equation}
The important point to stress is that, as the source of randomness in our description comes from the position of the nucleons, the joint probability density functions of $n$ variables, $\rho^{(n)}({\bf r}_1, \ldots, {\bf r}_n)$, are interpreted as the $n$-body densities of the nuclear ground state in the coordinate representation,
\begin{equation}
    \rho^{(n)}({\bf r}_1, \ldots, {\bf r}_n) = \int_{{\bf r}_{n+1}, \ldots, {\bf r}_A} |\langle {\bf r}_1, \ldots, {\bf r}_A | \Psi \rangle|^2 \,,
\end{equation}
which further implies
\begin{equation}
    \int_{{\bf r}_1, \ldots, {\bf r}_n} \rho^{(n)}({\bf r}_1, \ldots, {\bf r}_n) = 1.
\end{equation}
We see, then, that the local average of the entropy density, $\langle s({\bf x}) \rangle$, involves the expectation value of operators that involve only one random variable, ${\bf r}_i \equiv {\bf r}_1$. Therefore, Eq.~(\ref{eq:aveSappe}) becomes
\begin{align}
  C_1({\bf x}) \equiv  \langle s({\bf x}) \rangle =  s_0 \, A \int_{{\bf r}_1} \rho^{(1)}({\bf r}_1) g({\bf x}-{\bf r}_{\perp 1}) \,.
\end{align}

\subsubsection{Two-point function}

The two-point function of the entropy density is given by the average
\begin{align}
\nonumber   & \langle s({\bf x}) s({\bf y}) \rangle = \left \langle \biggl ( s_0 \frac{t({\bf x})+t^\prime ({\bf x})}{2}  \biggr )\biggl ( s_0 \frac{t({\bf y})+t^\prime({\bf y})}{2} \biggr) \right \rangle \\
\nonumber   & = \frac{s_0^2}{4} \biggl ( \bigl \langle t({\bf x})t({\bf y}) \rangle + \langle t({\bf x})t^\prime({\bf y}) \rangle  \\
   & \hspace{50pt } + \langle  t^\prime({\bf x})t({\bf y}) \rangle + \langle t^\prime({\bf y})t^\prime({\bf y}) \rangle  \biggr ).
\end{align}
Once again, as the two colliding nuclei are the same, and as all nucleons are treated equally, this expression simplifies to
\begin{align}
\nonumber    \langle s({\bf x}) s({\bf y}) \rangle &= \frac{s_0^2}{4} \bigl ( 2 \langle t({\bf x})t({\bf y}) \rangle + 2 \langle t({\bf x})t^\prime({\bf y}) \rangle \bigr) \\
\nonumber    &= \frac{s_0^2}{2} \biggl ( \sum_{i,j}^A \left \langle g({\bf x}-{\bf r}_{\perp i}) g({\bf y}-{\bf r}_{\perp j}) \right \rangle \\
\label{eq:S2appe}    &\hspace{20pt}+\sum_{i}^A \left \langle g({\bf x}-{\bf r}_{\perp i}) \right \rangle  \sum_{j}^A \left \langle g({\bf y}-{\bf r}_{\perp j}) \right \rangle  \biggr ) \,.
\end{align}
where we used $\left \langle t({\bf x}) t^\prime({\bf y}) \right \rangle = \langle t({\bf x}) \rangle \langle t({\bf y}) \rangle$, which is due to the fact that the positions of the nucleons in the first nucleus are independent of the positions of the nucleons in the second nucleus.

In the expression of the expectation value $\left \langle g({\bf x}-{\bf r}_{\perp i}) g({\bf y}-{\bf r}_{\perp j}) \right \rangle$, the subscripts $i$ and $j$ run from 1 to $A$, such that there are $A$ terms with $i=j$. These terms have to be separated, as they only involve the expectation value of a one-body operator. Therefore, we write
\begin{align}
\nonumber   & \sum_{i,j}^A  \left \langle g({\bf x}-{\bf r}_{\perp i}) g({\bf y}-{\bf r}_{\perp j}) \right \rangle =  A \langle g({\bf x}-{\bf r}_{\perp i}) g({\bf y}-{\bf r}_{\perp i}) \rangle \\
  & \hspace{70pt}+ (A^2-A)\langle g({\bf x}-{\bf r}_{\perp i}) g({\bf y}-{\bf r}_{\perp j}) \rangle \, ,
\end{align}
such that, with ${\bf r}_i \equiv {\bf r}_1$ and ${\bf r}_j \equiv {\bf r}_2$, Eq.~(\ref{eq:S2appe}) becomes
\begin{align}
\nonumber    &\langle s({\bf x}) s({\bf y}) \rangle = \frac{s_0^2}{2} \biggl [ A \int_{{\bf r}_1} \rho^{(1)} ({\bf r}_1) g({\bf x}-{\bf r}_{\perp 1}) g({\bf y}-{\bf r}_{\perp 1}) \\
\nonumber    & \hspace{20pt} + (A^2-A)  \int_{{\bf r}_1, {\bf r}_2} \rho^{(2)}({\bf r}_1, {\bf r}_2) g({\bf x}-{\bf r}_{\perp 1}) g({\bf y}-{\bf r}_{\perp 2}) \\
    & \hspace{20pt} + A^2 \int_{{\bf r}_1} \rho^{(1)} ({\bf r}_1) g({\bf x}-{\bf r}_{\perp 1})  \int_{{\bf r}_2} \rho^{(1)} ({\bf r}_2)   g({\bf y}-{\bf r}_{\perp 2})\biggr ] \,.
\end{align}
From this one obtains the connected two-point function according to
\begin{equation}
    C_2({\bf x}, {\bf y}) \equiv \langle s({\bf x}) s({\bf y}) \rangle - C_1({\bf x})C_1({\bf y}).
\end{equation}

\subsection{Derivation of the mean squared anisotropy}

We start from the leading-order formula for the mean squared anisotropy
\begin{equation}
 \langle \varepsilon_n^2 \rangle     =  \, \frac{ \int_{{\bf x}, {\bf y}} |{\bf x}|^n~|{\bf y}|^n~e^{in(\phi_x-\phi_y)} \, C_2({\bf x},{\bf y}) } { \left ( \int_{\bf x} |{\bf x}|^n \,C_1({ \bf x} )  \right )^{2} },
\end{equation}
and the correlation functions of the entropy density field given by
 \begin{align}
  \nonumber   C_1({\bf x}) &= s_0 \, A \int_{{\bf r}_{1}} \rho^{(1)}({\bf r}_{1}) g({\bf x}-{\bf r}_{1\perp}), \\
\nonumber C_2({\bf x}, {\bf y})&= \frac{s_0^2}{2} \biggl [  A \int_{{\bf r}_{1}} \rho^{(1)} ({\bf r}) g({\bf x}-{\bf r}_{1\perp}) g({\bf y}-{\bf r}_{1\perp})  \\ 
\nonumber &\hspace{-25pt} + (A^2-A) \int_{{\bf r}_{1}, {\bf r}_{2}} \rho^{(2)} ({\bf r}_{1}, {\bf r}_{2}) g({\bf x}-{\bf r}_{1\perp}) g({\bf y}-{\bf r}_{2\perp}) \\
& \hspace{-25pt} - A^2 \int_{{\bf r}_1} \rho^{(1)}({\bf r}_1) g({\bf x}-{\bf r}_{1\perp})\int_{{\bf r}_2} \rho^{(1)}({\bf r}_2) g({\bf y}-{\bf r}_{2\perp}) 
\biggr ] \, .
 \end{align}
Noting that the last term in the connected two-point function does not contribute to the mean squared eccentricity because of the spherical symmetry of the nuclear one-body density, one obtains
\begin{align}
\label{eq:msEnI}
\nonumber  &  \langle \varepsilon_n^2 \rangle = \frac{1}{2A^{2}} \frac{1}{\left ( \int_{{\bf r}_{1 }} \rho^{(1)} ({\bf r}_{1}) I_{n,0}({\bf r}_{1\perp }) \right )^2 } \\
 \nonumber   &  \biggl [ A \int_{{\bf r}_{1}} \rho^{(1)} ({\bf r}_{1  }) |I_{n,n}({\bf r}_{1\perp })|^2   \\
   & 
    \hspace{10pt} +  (A^2-A) \int_{{\bf r}_{1  },{\bf r}_{2  }} \rho^{(2)} ({\bf r}_{1 },{\bf r}_{2}) I_{n,n}({\bf r}_{1\perp })I^*_{n,n}({\bf r}_{2\perp })  \biggr],
\end{align}
where we introduce the following integrals
\begin{equation}
    I_{n,m}({\bf r}_\perp) \equiv \int_{\bf x} |{\bf x}|^n e^{im\phi_x} g({\bf x}-{\bf r}_\perp) \, .
\end{equation} 

In the (realistic) limit where the nucleonic profile, $g({\bf x})$, decays  much more sharply than the typical variation of ${\bf x}^n$, one can take a point-like approximation, $g({\bf x})=\delta({\bf x})$,  which results in
\begin{align}
\label{eq:I0}
    I_{n,0} &\approx r_\perp^n, \\
    \label{eq:In}
   I_{n,n} ({\bf r}_\perp) &\approx (r_{x}+ir_{ y})^n \,, 
\end{align}
where we use ${\bf r}_\perp=(r_{ x}, r_{y})$, $r_\perp=|{\bf r}_\perp|$. These results lead, thus, to the expressions in Eq.~(\ref{eq:Enquasifinal}) for the mean squared eccentricities.

It is interesting to note that in the commonly-adopted scenario of a normalized Gaussian nucleon profile, 
\begin{equation}
    g({\bf x}) = \frac{1}{2\pi w^2} e^{-\frac{|{\bf x}|^2}{2w^2}} \, ,
\end{equation}
the integral in Eq.~(\ref{eq:In}) turs out to be exact, that is, independent of the choice of $w$. The other integrals with $m=0$ give instead
\begin{align}
  \nonumber   I_{2,0} ({\bf r}_\perp) &= 2w^2 + r_\perp^2, \\
  \nonumber  I_{3,0}({\bf r}_\perp) &= \frac{\sqrt{\pi}}{\sqrt{8} w} e^{-\frac{r_\perp^2}{4w^2}} \biggl [ \left (6w^4 + 6 w^2 r_\perp^2 + r_\perp^4  \right ) I_0 \left ( \frac{r_\perp^2}{4w^2} \right ) \\
    &\hspace{60pt} + r_\perp^2 (4w^2 + r_\perp^2) I_1\left ( \frac{ r_\perp^2 } {4 w^2} \right ) \biggr]. 
\end{align}
As anticipated, these formulas yield Eq.~(\ref{eq:I0}) in the limit $w\rightarrow0$.

\subsection{Intrinsic Gaussian density}

Our starting point is an intrinsic nuclear shape given by a (normalized) Gaussian profile of the form 
\begin{equation}
    \rho(r,\Omega) \equiv \frac{1}{(2\pi)^{3/2} R^3} \exp \left ( -\frac{r^2}{2 R(\Omega)^2}\right ) \, ,
\end{equation}
where $\Omega$ denotes the set of Euler angles stipulating the orientation of the intrinsic deformation in space and where the nuclear surface is expanded in terms of axial quadrupole and octupole deformations according to
\begin{equation}
    R (\Omega) \equiv R \left [ 1 + \beta_{20} \ylm{2}{0}{(\Omega)}  + \beta_{30} \ylm{3}{0}{(\Omega)}    \right ] \, .
\end{equation}
Performing a Taylor expansion to first order in powers of the deformation parameters yields
\begin{equation}
    \rho(r, \Omega) \approx \frac{e^{-r^2 / (2R)^2}}{(2\pi)^{3/2} R^3} \biggl [ 1 + \frac{ r^2 \ylm{2}{0}{(\Omega)}}{R^2} \beta_{20} + \frac{ r^2 \ylm{3}{0}{(\Omega)} } {R^2} \beta_{30} \biggr ] \, .
\end{equation}
Rotating the intrinsic density in all possible ways, the lab-frame one-body density of the $J=0$ state is given by the isotropic part of the former
\begin{equation}
    \rho^{(1)}({\bf r}_1) \equiv \frac{1}{4\pi} \int_\Omega  \rho_\Omega({\bf r}_1) =  \frac{1}{(2\pi)^{3/2} R^3} \exp \left ( -\frac{r^2}{2 R^2}\right ) \, .
\end{equation}
Correlations originate from the combination of the preferred orientation of the, otherwise independent, nucleons in the intrinsic frame and of their collective rotation. As a result, the lab-frame two-body density of the $J=0$ ground state is obtained through
\begin{equation}
    \rho^{(2)}({\bf r}_1, {\bf r}_2) \equiv \frac{1}{4\pi} \int_\Omega  \rho_\Omega({\bf r}_1) \rho_\Omega({\bf r}_2) ,
\end{equation}
leading to a long expression that is not displayed here for brevity.

Based on these quantities, the expectation values appearing in Eq.~\eqref{eq:Enquasifinal} for $n=2,3$ are given by
\begin{subequations}
\begin{align}
\label{eq:R2}    \langle \hat{R}_2 \rangle &=  2 R^2 \, , \\
\label{eq:R3}       \langle \hat{R}_3 \rangle &=  \sqrt{ \frac{9\pi}{2} } R^3 \, , \\
     \langle \hat{R}_4 \rangle &=  8 R^4 \, , \\
        \langle \hat{R}_6 \rangle &=  48 R^6 \, , \\   
 \langle \hat{\mathcal{E}}_2 \rangle &= \frac{6}{\pi} R^4 \,\beta_{20}^2 \, , \\
\langle \hat{\mathcal{E}}_3 \rangle &= \frac{18432}{245\pi^2} R^6 \, \beta_{30}^2 \, .
\end{align}
\end{subequations}
Putting all these results together into Eq.~(\ref{eq:Enquasifinal}), one arrives at the expressions of the mean squared anisotropies shown in Eq.~(\ref{eq:Engauss})

\subsection{Numerical validation}

The leading-order picture of heavy-ion collisions is based on the following four simplifications:
\begin{enumerate}[label=\textbf{\arabic*}.]
    \item All nucleons in the colliding ions participate in the interaction (or $N_{\rm part}=2A$).
    \item The entropy density is obtained from the arithmetic average of two thickness functions.
    \item Collisions occur at zero impact parameter.
        \item We truncate the expression of $\left \langle \varepsilon_2^2 \right \rangle$ to the leading order in the fluctuations, that is, we use Eq.~(\ref{eq:linEn}).
\end{enumerate}
The goal in this section is to show that the conclusions of the analysis presented in the bulk of the paper hold even if all the mentioned approximations are lifted. More precisely, we want to demonstrate in a more realistic treatment of the collisions that indeed the effect of the two-body operator $\hat{\mathcal{E}}_n$ captures the enhancement of the mean squared ellipticity, $\langle \varepsilon_2^2 \rangle$, the latter being modeled to originate from the intrinsic nuclear deformation.

To do so, different options for the deposition of the entropy density are considered, \textit{i.e.} we use the convenient \trento{} Ansatz according to which
    \begin{equation}
\frac{dS}{dy}~[1/{\rm fm}^2] \equiv s({\bf x}) = \left ( \frac{ t({\bf x})^p + t^\prime({\bf x})^p } {2} \right)^{1/p} \, ,
\end{equation}

First, to remove the simplification \textbf{1} above, the thickness functions $t({\bf x})$ and $t^\prime({\bf x})$ are constructed from the superposition of the \textit{participant} nucleons, as determined by the \trento{} code, instead of all nucleons. Second, to remove simplification \textbf{2} three representative models are used \cite{Moreland:2014oya}: 
\begin{itemize}
    \item A harmonic mean model:
\begin{equation}
\nonumber  p=-1~~ \longrightarrow~~  s({\bf x}) \propto  \frac{t({\bf x})t'({\bf x})}{t({\bf x})+t'(\bf x)}. 
\end{equation}
    \item A geometric mean model:
\begin{equation}
 \nonumber  p=0~~ \longrightarrow~~   s({\bf x}) \propto  \sqrt{t({\bf x})t'({\bf x})}. 
\end{equation}
    \item An arithmetic mean model:
\begin{equation}
\nonumber  p=1 ~~\longrightarrow ~~  s({\bf x}) \propto  \frac{t({\bf x})+t'({\bf x})}{2}. 
\end{equation}
\end{itemize}
In the limit where $t({\bf x})=t'({\bf x})$, the above prescriptions return the same result, \textit{i.e.} $s({\bf x}) \propto t({\bf x})$. It is reasonable to expect that in the limit of ultra-central collisions, where most of the nucleons are involved in the interaction, the results of the simulations depend little on the specific choice of $p$ (see also Ref.~\cite{Jia:2025wey}).

\begin{figure}[t]
    \centering
    \includegraphics[width=\linewidth]{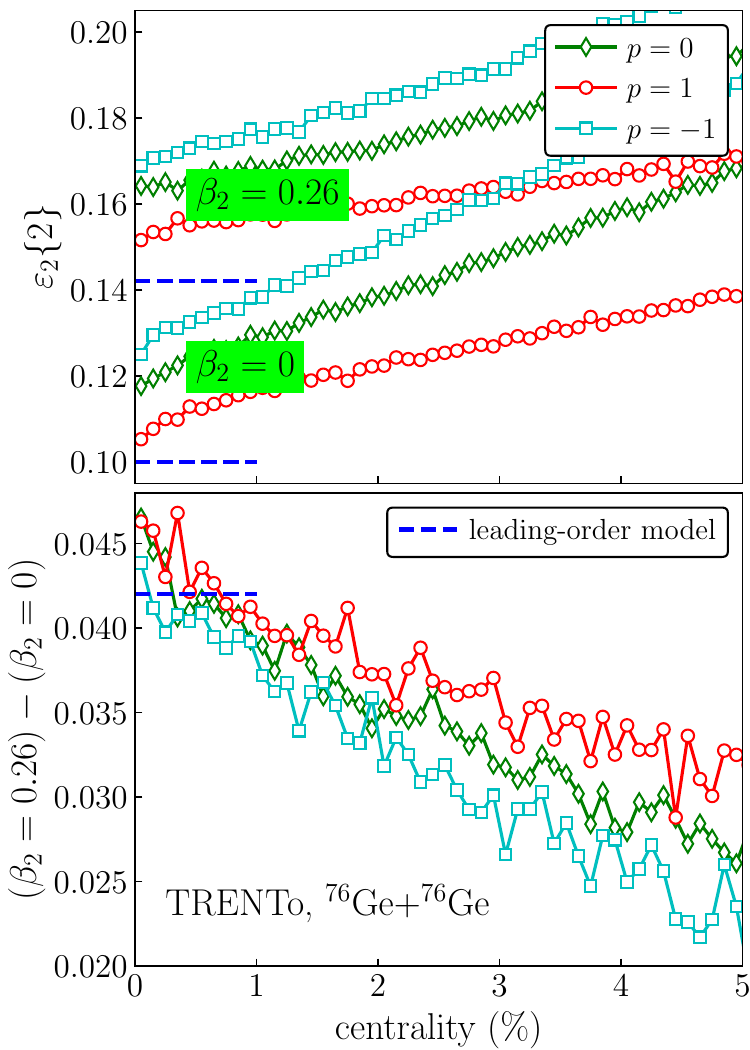}
    \caption{\textit{Top:} Values of $\varepsilon_2\{2\}$ as a function of the centrality percentile in \gege{} collisions. Different symbols correspond to different \trento{} parametrizations for the entropy density, as discussed in the text. The results are obtained for both spherical ($\beta_2=0$) and deformed ($\beta_2=0.26$) ions. \textit{Bottom:} We show the difference between the value of $\varepsilon_2\{2\}$ obtained in collisions of deformed nuclei and the same quantity obtained for spherical nuclei. In both panels, the dashed blue line indicates the prediction of the leading-order model, where Eq.~(\ref{eq:linEn}) is evaluated in Monte Carlo simulations with an initial entropy density defined as in Eq.~(\ref{eq:save}).}
    \label{fig:e22}
\end{figure}

We collide $^{76}$Ge nuclei. This isotope is one of the candidates for neutrinoless double beta decay \cite{Agostini:2022zub}, which makes an analysis of its structure particularly compelling. The nucleonic density (both protons and neutron) is assumed to be given by a Woods-Saxon distribution with a sole axial quadrupole deformation  [${\bf r}=(r, \Omega)$], \textit{i.e.}
\begin{equation}
\label{eq:Ws}
    \rho ({\bf r}) \propto \frac{1}{1+e^{\frac{r-R(\Omega)}{a}}},
\end{equation}
with
\begin{equation}
    R(\theta) \equiv R_0 [1 + \beta_2 \ylm{2}{0}{(\Omega)}].
\end{equation}
The parameters are taken to be $a=0.52$ fm and $R=4.60$ fm along with two possible values for the quadrupole deformation $\beta_2=0$ or $\beta_2=0.26$ \cite{Grams:2023sml}. The triaxiality in this nucleus, discussed at length in Ref.~\cite{Rodriguez:2017afo,Ayangeakaa:2019psv,Belley:2023lec} is presently omitted, as it does not contribute to leading order to the mean squared $\varepsilon_n$ values \cite{Jia:2021tzt}. Once the thickness functions are sampled, the entropy density profiles are generated following the \trento{} prescriptions discussed above, for $p=-1$, $p=0$, and $p=1$. For each model, $10^7$ collisions are generated and the events are sorted in \textit{centrality} classes defined according to their total entropy. In doing so, the approximation \textbf{3} is effectively lifted given that a large spread of the values of the impact parameters is allowed even in central events.

In each event, the initial spatial anisotropy, $\mathcal{E}_n$, is evaluated according to Eq.~(\ref{eq:newEn}). Eventually, simplification \textbf{4} is also removed by evaluating the mean squared anisotropy by averaging over events instead of taking the perturbative formula. Specifically, the root mean squared anisotropy 
\begin{equation}
    \varepsilon_2\{2\} \equiv \sqrt{ \langle \varepsilon_2^2 \rangle },
\end{equation}
is computed as a function of the collision centrality, where the limit 0\% corresponds to events with the highest total entropy, $S$, \textit{i.e.} ultra-central collisions.

The results are shown as symbols in Fig.~\ref{fig:e22}. In the upper panel, the mean squared ellipticity is seen to receive a positive correction from the deformation of the nucleus, which enhances the fluctuations of the ellipticity of the overlap area. The three models lead to similar numerical values for the ellipticity toward the limit of 0\% centrality, in agreement with the expectation of the \trento{} formula. The blue dashed line in Fig.~\ref{fig:e22} represents the result of the leading-order model, where we evaluate numerically Eq.~(\ref{eq:linEn}) using the entropy density in Eq.~(\ref{eq:save}).

Although the models show a significant deviation from the leading-order formula, they all seem to be affected by the inclusion of $\beta_2=0.26$ in a similar way. In the lower panel, the difference between $\varepsilon_2\{2\}$ obtained for deformed  ($\beta_2=0.26$) and spherical ($\beta_2=0$) nuclei is displayed. Remarkably, the results of the three models collapse onto the same value in the limit of central collisions. This value is perfectly consistent with the prediction of the leading-order model, according to which ($A\gg1$)
\begin{equation}
    \varepsilon_2\{2\}_{\beta_2 = 0.26}^2 - \varepsilon_2\{2\}^2_{\beta_2=0} = \frac{\langle \hat{\mathcal{E}}_n \rangle}{2 \langle \hat{R}_2 \rangle^2}.
\end{equation}
Therefore, this formula appears to capture the correction induced by the presence of a deformation parameter irrespective of the detailed model features present in the \trento{} simulations.

\subsection{Radii and eccentricities in spherical coordinates}

\paragraph{Transverse radii.} We want to give a definition of the transverse radius in spherical coordinates. The quantity  we are interested in reads
\begin{equation}
    r_\perp^{n} = (r^2 - z^2)^{n/2} = (r^{2} - r^2 \cos^2 \theta)^{n/2} = r^{n}\, \sin^{n} \theta.
\end{equation}
For even powers of the radius, we use the definition of the spherical harmonics in their maximal projections,
\begin{equation}
    \ylm{n}{\pm n}{(\Omega)} = c_{\pm n} e^{\pm i n \phi} \sin^n \theta, 
\end{equation}
from which we can conveniently write
\begin{equation}
    r_\perp^{2n} = (c_n c_{-n})^{-1} \,  r^{2n} \, \ylm{n}{n}{(\Omega)} \, \ylm{n}{-n}{(\Omega)} \, .
    \end{equation}
To proceed we use the following property of the spherical harmonics,
\begin{align}
    \ylm{l}{m}{(\Omega)} \, &\ylm{l'}{m'}{(\Omega)} =   \notag \\
    & \sum_{\lambda \mu} (-1)^\mu 
    \sqrt{\frac{\jhat{l} \jhat{l'} \jhat{\lambda} }{4 \pi}}
    \threej{l}{m}{l'}{m'}{\lambda}{-\mu}
    \threej{l}{0}{l'}{0}{\lambda}{0}
    \ylm{\lambda}{\mu}{(\Omega)} \, ,
\end{align}
where we introduced the short-hand notation $\hat{x}\equiv \sqrt{2x+1}$.
From the expansion of the spherical harmonics we get an expression for the required products of the form
\begin{align}
    \ylm{n}{n}{(\Omega)} \, \ylm{n}{-n}{(\Omega)} = 
    \frac{(-1)^n}{\sqrt{4\pi}} \ylm{0}{0}{(\Omega)}
    + \sum_{k=1}^n d_k \, \ylm{2k}{0}{(\Omega)}
\end{align}
for some expansion coefficients $d_k$ involving higher spherical harmonics $\ylm{2k}{0}{}$.
Since $\ylm{0}{0}{(\Omega)} = 1/\sqrt{4\pi}$ and assuming that the many-body state has vanishing multipole moment $\langle Q_{lm} \rangle =0 $, the moments of the transverse radii are given by
\begin{subequations}
\begin{align}    
    \langle r_\perp^2 \rangle &= \frac{2}{3} \langle r^2 \rangle \, , \\ 
    \langle r_\perp^4 \rangle &= \frac{8}{15} \langle r^4 \rangle\, , \\
    \langle r_\perp^6 \rangle&= \frac{16}{35} \langle r^6 \rangle\, , 
\end{align}
\end{subequations}
It can be shown that all even moments of the transverse radius can be related to the normalization of the spherical harmonics to give a clean connection to the matter radius
\begin{align}
    \langle r_\perp^{2n} \rangle = \frac{|c_n|^2}{\sqrt{4 \pi}} \langle r^{2n} \rangle \, .
\end{align}

Now, for the evaluation of the spatial triangularity, or $\langle \varepsilon_3^2 \rangle$ in Eq.~(\ref{eq:Enquasifinal}), we have introduced a moment $\langle r_\perp^3 \rangle^2$ in the denominator. Because of the odd exponent, this does not have a simple decomposition in terms of spherical harmonics. One possibility is to redefine $\varepsilon_n$ with the cube of the mean squared radius in the denominator, $\langle r^2 \rangle^3$, as recently suggested in Ref.~\cite{Sousa:2024msh}. In either case, the following relation can be used. For large enough nuclei, the one-body density is typically close to a uniform sphere with a sharp surface. In that limit, one obtains:
\begin{equation}
 \left \langle r_\perp^2 \right \rangle^3 = 1.48  \, \left  \langle r_\perp^3 \right \rangle ^2 .
\end{equation}
We have checked that this relation holds, up to percent-level corrections, in the case of a realistic Woods-Saxon density such as that shown in Eq.~(\ref{eq:Ws}). Therefore, it provides us with a convenient way to calculate the denominator of the mean squared triangularity, $\left \langle \varepsilon_3^2 \right \rangle$.

\paragraph{Eccentricity operators.}

The evaluation of the \textit{eccentricity} operator relies on its representation in spherical coordinates
\begin{align}
    \hat{\mathcal{E}}_n(\mathbf{r}_1,\mathbf{r}_2) &= |c_n|^{-2} \,  r_1^n \,  \ylm{n}{n}{(\Omega_1)} \, r_2^n \, \ylm{n}{-n}{(\Omega_2)} \\
    &= \hat{F}_{n} ({\bf r}_1) \hat{F}_{-n} ({\bf r}_2) \, ,
\end{align}

Many-body calculations typically employ the second-quantized formalism where the operators are represented in a one-body basis of, \textit{e.g.} the spherical Harmonic Oscillator,
\begin{align}
    |p\rangle \equiv c^\dagger_p | 0 \rangle \, ,
\end{align}
where $|0\rangle$ denotes the physical vacuum and $c^\dagger_p$  the single-particle creation operator that generates the state $|p\rangle$.
As the eccentricity operator is a product of two one-body operators, its second-quantized representation involves one- and two-body parts 
\begin{align}
    \hat{\mathcal{E}}_n  = 
    \sum_{pq} \bar \epsilon^{(n)}_{pq} c^\dagger_p c_q +
    \frac{1}{4} \sum_{pqrs} \bar \epsilon^{(n)}_{pqrs} c^\dagger_p c^\dagger_q c_s c_r \, ,
\end{align}
whose matrix elements are given by
\begin{subequations}
\begin{align}
    \bar \epsilon^{(n)}_{pq} &= \sum_r f^{(n)}_{pr} f^{(-n)}_{rq} \, ,   \\
    \bar \epsilon^{(n)}_{pqrs} &= 2 \Big(f^{(n)}_{pr} f^{(-n)}_{qs} - f^{(-n)}_{ps} f^{(n)}_{qr} \Big) \, .
\end{align}
\end{subequations}
Here the short-hand notation for the matrix elements, \textit{e.g.}, $f^{(\pm n)}_{pq} \equiv \langle p | \hat F_{\pm n} | q \rangle$ was employed. 
Further notice that two-body matrix elements are explicitly anti-symmetrized, \textit{i.e.}, $\bar \epsilon_{pqrs} = - \bar \epsilon_{qprs}$.

\bibliographystyle{apsrev4-1}
\bibliography{bibb}

\end{document}